\documentclass[prl,aps,showpacs]{revtex4}
\usepackage{times}
\usepackage{amsmath}
\usepackage{amsfonts,amssymb}
\usepackage{verbatim}
\allowdisplaybreaks[4]

\DeclareMathOperator{\imag}{Im}
\begin{document}

\title{New mechanism of pulsar radio emission}

\author{M. Gedalin and E. Gruman}
\affiliation{Department of Physics,
Ben-Gurion University, Beer-Sheva 84105, Israel}
\author{D.B. Melrose}
\affiliation{School of  Physics,
University of Sydney, NSW 2006, Australia}
\date{\today}

\begin{abstract}
{ It is  shown that pulsar radio emission can be generated
effectively through a streaming motion in the polar-cap regions of a pulsar
magnetosphere causing nonresonant growth of waves that can escape directly.
As in other beam models, a relatively low-energy high-density beam is
required. The instability generates quasi-transverse waves in a beam mode at
frequencies that can be well below the resonant frequency. As the waves
propagate outward growth continues until the height at which the wave
frequency is equal to the resonant frequency.  Beyond this point the 
waves  escape
in  a natural plasma mode (L-O mode).
This one-step  mechanism is much more efficient than
previously widely considered multi-step mechanisms.}
\end{abstract}

\maketitle

Since the discovery of pulsars in 1967 the mechanism of generation of
their highly nonthermal (brightness temperatures up to $10^{29}\,$K)
pulsed radio emission (in the range $10^8-10^{11}\,$Hz) remains one of the
most intriguing astrophysical puzzles \cite{MT77}. Compact sizes (radius
$R_p\sim 10^6\,$cm), fast rotation (period  $P\sim1\,$s), and
superstrong polar magnetic fields ($B\sim 10^{12}\,$G) result in the
efficient avalanche production of an ultrarelativistic pair plasma (Lorentz
factor $\gamma_p\gg 1 $) in the vicinity of the { magnetic poles of the
neutron star} (see, e.g. Ref.~\onlinecite{Arons83}).  This plasma flows
outward along the open magnetic field lines and escapes the pulsar
magnetosphere as a relativistic wind beyond the light cylinder, $R=cP/2\pi$.
In a standard polar-cap model, a primary beam (Lorentz factor $\gamma_b\gg
\gamma_p$) of particles of one species propagates through a secondary pair
plasma \cite{MT77}. The superstrong magnetic field of the pulsar implies a
very short lifetime for the electrons and positrons to radiate away all
their perpendicular momenta, so that the plasma distribution is one
dimensional. The properties of this pulsar plasma determine the natural wave
modes, and the problem is to explain how excitation of these modes occurs
and how it produces the observed radiation that escapes from the
magnetosphere.

It is probable that the radio spectrum forms in the  inner
magnetosphere \cite{Cordes92}, where the infinite magnetic field
approximation is appropriate. The properties of low-frequency (well below
the cyclotron frequency) waves in a one-dimensional, relativistic pair
plasma have been extensively studied (e.g., Ref.~\onlinecite{AR00} and
references therein). It has been found \cite{Gedalin98,MG99} that for a
rather wide class of plasma distributions the natural modes are the
electromagnetic $t$-mode, $\omega=kc$,  and two mixed (with transverse and
longitudinal components of the electric field vector) modes, the almost
nondispersive Alfv\'{e}n mode, $\omega=k_\parallel v_A$ (here and hereafter
subscripts $\parallel$ and $\perp$ refer to the direction with  respect to
the external magnetic field), and the L-O mode which has a long wavelength
cutoff. The L-O mode is superluminal, $\omega/k_\parallel >c$, near the
cutoff but can be subluminal for sufficiently small angles, $\theta$, of
propagation at high frequencies. The critical features of any plausible
mechanism for the radio wave generation can be summarized as follows
\cite{MG99}. Only $t$-waves (which cannot be generated through a beam
instability) and waves which eventually appear on the L-O branch can freely
leave the pulsar magnetosphere. Beam instabilities are widely favored, and in
the pulsar magnetosphere are of hydrodynamical type, where the whole beam
excites the modes, in contrast to  kinetic instabilities which are driven
only by a group of resonant particles
\cite{M91}.

{  One of the most  widely favored scenarios for
the radio emission mechanism involves a resonant instability in which an
energetic  beam causes quasi-longitudinal subluminal waves to grow. This
mechanism encounters  certain difficulties.
The first difficulty is that the emission mechanism is {\it
indirect}: the postulated growing waves need to be converted into
quasi-transverse waves (in the $t$ or L-O modes) via some  secondary (e.g.,
nonlinear) mechanism before they can produce escaping radiation. A {\it
direct\/} mechanism in which the escaping waves are generated directly by
the instability would be preferable, and the mechanism proposed here has
this desirable feature. The second difficulty concerns the beam: when the
generation is attributed to the primary, highly  relativistic,
$\gamma_b\sim 10^7$, the growth is too slow to be effective. Thus,
efficient beam instability
requires a denser, lower
energy beam. One suggestion is that such a beam results from nonstationary
avalanche pair generation \citep{Usov87,AM98}. Alternatively, even in the
steady avalanche regime, the high energy  tail of the pair plasma
distribution \citep{DH82} transforms into a  dense beam due to the inverse
Compton scattering \citep{Lyu00}. The resulting  distribution typically
consists of a plasma body with $\gamma_p\sim 10^1-10^2$  and a beam with
$\gamma_b\sim 10^3$. In the steady case such a beam forms  at about 
$10R_p$. Although existence of such a beam seems plausible it is still 
a model dependent assumption. 
The third difficulty is related to the  nature of the wave
generation.  Most authors have concentrated on
resonant excitation of parallel (to the magnetic field) propagating
modes (Langmuir waves)
\cite{Magneville90,Ly98,AR00,Lyu00}, where it is widely believed that there
is a sharp maximum of the growth rate. Growth for slightly oblique
propagation has been considered \cite{Asseo90,MG99} and also found to favor
quasi-longitudinal L-O waves.  Only subluminal waves can be
resonantly excited by a beam. L-O mode becomes subluminal only at frequencies
$\sim\gamma_p\times$plasma frequency \cite{MG99}.
  Growth is possible at the
resonant frequency, which is just above the frequency where the waves become
subluminal. These  L-O waves  can eventually escape directly. However,
the  frequencies are too high to
account for the broadband pulsar emission extending to much lower
frequencies (so that a conversion mechanism to lower frequencies would be
required). Moreover, growth is restricted
to a narrow range of heights in the magnetosphere where the resonance
condition is satisfied, restricting the growth factor to too small a value to
allow effective growth \cite{MG99}. These constraints led to our relatively
pessimist view in \cite{MG99} of the effectiveness of direct growth of
escaping quasi-transverse waves. }

{ In this Letter we show that a nonresonant version of
the instability causes a beam mode to grow over a broad range of lower
frequencies. These waves  can grow over a large range
of heights. As the propagate outward that ratio of their frequency to the
resonant frequency decreases, and at which height at which this ratio become
unity, they evlove into L-O mode waves, and escape, without any secondary
conversion process being required. }

We use  the following notation:  $p=mu$ is the (one dimensional) particle
momentum, with $m$ the electron mass, and  $v=u/\gamma$,
$\gamma=(1-v^2)^{-1/2}=(1+u^2)^{1/2}$ is the particle velocity in units with
$c=1$. We assume that the distribution function consists of a pair ($p$) and
a beam ($b$) component,
$f(u)=n_pf_p(u)+n_bf_b(u)$ with $\int f_p du=\int f_b du=1$.  We adopt the
infinite magnetic field limit, which is justified in the inner part of the
plasma magnetosphere where the beam-plasma system is formed \cite{Lyu00},
and then it is not necessary to distinguish between electrons and positrons,
which contribute in an identical manner. The dispersion relation for this
beam-plasma system takes the  following form \cite{Gedalin98,MG99,AR00}:
\begin{equation}
     \epsilon_\parallel =\frac{\tan^2\theta}{z^2-1},
      \label{eq:dr}
\end{equation}
with $z =\omega/ k_\parallel$,
$k_\parallel=k\cos\theta$, $k_\perp=k\sin\theta$,  and where
$\epsilon_\parallel=1-(\omega_{p}^2/k_\parallel^2)W(z)+\epsilon_b$
is the parallel dielectric constant. Here $\omega_p=(8\pi n_pe^2/m)^{1/2}$
(equal electron and positron densities) is the plasma frequency for the
pairs, $W=\int_{-\infty}^\infty(v-z-i\tau)^{-1} (df_p/du)du$,
$\tau\rightarrow +0$, is the dispersion function for the pairs; the
contribution of the beam is
$\epsilon_b=-\omega_b^2\gamma_b^{-3}(\omega-k_\parallel
v_b)^{-2}$. { We consider a cold beam for simplicity; the cold
approximation is justified when the instabilities are know to be
hydrodynamic. } The dispersion relation
\eqref{eq:dr} takes the form
\begin{equation}
\frac{\omega_b^2}{\gamma_b^3(\omega-k_\parallel v_b)^2}=
\frac{z^2-\cos^{-2}\theta}{z^2-1}-\frac{\omega_p^2W(z)}{k_\parallel^2}\equiv
K(z).
      \label{eq:dr1}
\end{equation}
In the absence of the beam the dispersion relation for the natural modes of
the pair plasma is $K(z)=0$, which can be written in the parametric form
(recall $c=1$)
\begin{equation}
k_\parallel^2=\frac{\omega_p^2W(z^2-1)}{z^2-\cos^{-2}\theta},
\quad  \omega^2= \frac{\omega_p^2Wz^2(z^2-1)}{z^2-\cos^{-2}\theta}.
\label{eq:omega}
\end{equation}
The inclusion of the beam introduces additional solutions, called beam
modes, and hydrodynamic instabilities may be attributed to a beam mode
becoming intrinsically growing. The instability is said to be nonresonant
when the beam mode does not coincide with a natural mode of the pair plasma,
and resonant when it does. The contribution of the beam is significant only
when the denominator in the  left hand side of \eqref{eq:dr1} is small, that
is, near $z=v_b$. Writing $\omega=k_\parallel v_b+\delta \omega$,
$|\delta\omega|\ll|\omega|$ one finds
\begin{equation}
      \delta \omega=\omega_b\gamma_b^{-3/2}K(v_b)^{-1/2}=
      \omega_b\gamma_b^{-3/2}
      \left[\frac{v_b^2-\cos^{-2}\theta}{
   v_b^2-1} -\frac{\omega_p^2
      W(v_b)v_b^2}{\omega^2}\right]^{-1/2}.
      \label{eq:solution}
\end{equation}
The unstable solution  $\delta\omega=i\Gamma$, $\Gamma>0$, exists for
$K(v_b) <0$, that is,
$\omega^2<\omega_p^2W(v_b) v_b^2(v_b^2-1)/(v_b^2-\cos^{2}\theta)$, which
requires  $W(v_b)>0$. This is the nonresonant beam instability which sets on
the beam mode $\omega\approx k_\parallel v_b$.

The expression \eqref{eq:solution} becomes invalid for $K(v_b)\to 0$,
that is, where the beam mode $\omega=k_\parallel v_b$ resonates with the L-O
mode (which is the solution of $K(v_b)=0$). In this
case the right hand side of \eqref{eq:dr1} should be Taylor expanded
up to the first nonzero term which immediately gives (see, e.g.,
Ref.~\onlinecite{AR00})
$\Gamma_r=\imag
\delta\omega=(\sqrt{3}/2)(\omega_b^2/\gamma_b^3K')^{1/3}$, with
$K'\equiv(\partial K/\partial \omega)_{\text{res}}=
2v_b^2\gamma_b^4\tan^2\theta/\omega -
\omega_p^2v_b^3W'(v_b)/\omega^3$, and $W'(z)=dW/dz$. For a wide class of
distributions the approximation
$W'(v_b) \sim \gamma_p^2 W(v_b)$ holds. This implies that the resonant growth
rate is insensitive to $\theta \lesssim\gamma_p/\gamma_b^2$, and decreases
slowly with $\theta \gtrsim\gamma_p/\gamma_b^2$. { This point is
important in the following discussion: if the growth rate were very
sensitive to $\theta$, then a small change in $\theta$ as the waves
propagate outward along the curved field lines would restrict the
distance over which growth can occur, and hence severely limit the possible
growth factor.}

The polarization of the unstable mode is  given by
$(E_\perp/E_\parallel)=\tan\theta(1-z^2)^{-1}=\gamma_b^2\tan\theta$.
Thus, the unstable mode is quasi-longitudinal ($\mathbf{E}\parallel
\mathbf{B}_0$) for $\theta\lesssim \gamma_b^{-2}$ and quasi-transverse
($\mathbf{E}\perp\mathbf{B}_0$) otherwise. { The polarization of the
growing waves is relatively unimportant in practice: the polarization
evolves as the waves propagate outward and the observed polarization may be
quite different from the polarization at the point of emission
\citep{BA86}.}

The derived expressions give the growth rate for arbitrary propagation
angle and beam parameters. For small $\theta\ll 1$ and a highly relativistic
beam $v_b\approx 1-1/2\gamma_b^2$, $\gamma_b\gg 1$, the growth rate for
the nonresonant instability \eqref{eq:solution} reduces to
\begin{equation}
\Gamma_n=\omega_b\gamma_b^{-1/2} \left[ \frac{\omega_p^2
W(v_b)}{\omega^2} -1 -\gamma_b^2\theta^2\right]^{-1/2},
\label{eq:nressmall}
\end{equation}
which is a slightly increasing function of $\theta$. In the
range $\gamma_b^{-2} \lesssim \theta\lesssim \gamma_b^{-1}$, where the waves
are quasi-transverse, the dependence on $\theta$ is negligible. The
maximum unstable frequency
$\omega_m^2=\omega_p^2W(v_b)/(1+\gamma_b^2\theta^2)$,
is also the frequency at which the resonant instability occurs, and it
is also almost independent of $\theta$ in this range. We conclude that
under quite general conditions the beam instability generates weakly
oblique, quasi-transverse waves with a similar efficiency to that for
quasi-longitudinal waves.

{ To illustrate the results in the simplest possible way, we use the cold
plasma  approximation, $f_p=\delta(u-u_p)$.} In this case one has
$W(z)=\gamma_p^{-3}(z-v_p)^{-2}$. The parametric equations \eqref{eq:omega}
become
$k_\parallel^2=\omega_p^2(z^2-1)/\gamma_p^3(z^2 -\cos^{-2}\theta) (z-v_p)^2$,
$\omega =k_\parallel z$. For small $\theta$ and $\gamma_b\gg\gamma_p\gg1$,
the growth rate of the nonresonant instability becomes
   \begin{equation}
   \Gamma_n=\omega_b\gamma_b^{-3/2}
\left[4\omega_p^2\gamma_p/\omega^2
-1-\gamma_b^2\theta^2\right]^{-1/2}.
\label{eq:drcold1}
\end{equation}
Thus the growth rate for the resonant instability also decreases
monotonically with increasing $\theta$. { Although these results are
derived using the cold plasma expressions, as we plan to show in detail
elsewhere, they are illustrative of a rather wide class of distributions
due to the fact that the hydrodynamic instability is insensitive to the
details of the beam distribution.}

For small propagation angles $\theta\lesssim 1/\gamma_b$ the growth rates
are almost independent of $\theta$, and we can use the following
approximations. In the ultrarelativistic limit $\gamma_b\gg \gamma_p\gg 1$
one has
$\Gamma_n=\omega_b\gamma_b^{-3/2}(4\omega_p^2\gamma_p/\omega^2-1)^{-1/2}$.
For low frequencies, $\omega\ll 2\omega_p\gamma_p^{1/2}$, this expression
simplifies to $\Gamma_n=\omega\omega_b/2\omega_p\gamma_b^{3/2}\gamma_p^{1/2}$.
The resonant frequency is $\omega_r=2\omega_p\gamma_p^{1/2}$, and the
resonant growth rate is
$\Gamma_r=3^{1/2}2^{-4/3}(\omega_p\omega_b^2)^{1/3}/\gamma_b\gamma_p^{1/2}$.
Note that
$\Gamma_n(\omega=\omega_r)/\Gamma_r \sim (\gamma_p/\gamma_b)^{1/2}
(n_b/n_p)^{1/3}$ implies that for moderate $n_p\gamma_p/n_b\gamma_b$ the
ratio of the two growth rates is of the order of  unity. One can approximate
the growth rate in the whole range  by
$\Gamma=(\omega/2\gamma_b^{3/2}\gamma_p^{1/2}) H(\omega_r-\omega)$,
where $H(x\geq 0)=1$ and $H(x<0)=0$. This approximation is also valid if
the above ratio is small, except in a narrow frequency range around
that resonant frequency. We exploit this approximation in our estimates
below. { Note that the condition $\gamma_b\gg\gamma_p$ is made for
simplicity, and it is not an essential condition for the instability to
operate.}

The direct excitation of quasi-transverse waves is a fast process, faster
than any nonlinear conversion mechanism. Let the growth rate be
$\Gamma(\omega)$, which is a function of the plasma parameters, $n_p$, $n_b$,
$\gamma_p$, $\gamma_b$, and $n_p$, { through which it} depends on the
radius, $R$ from the center of the star. The wave amplitude evolves
according to $(da_\omega/dt)=\Gamma(\omega,R)a_\omega$. With the plasma
streaming outward at close to the speed of light, the solution implies
$a_\omega(R)=a_\omega(R_0)\exp(\int_{R_0}^R \Gamma(\omega,R) dR)$, where
$R_0$ is the radius where the instability sets in. The power spectrum of the
escaping radiation is proportional to the square of this amplitude.

In a homogeneous static plasma  the fastest growing mode is the resonant
one, and one might expect that the resonant frequency ultimately dominates
the spectrum. { In the inhomogeneous plasma of the pulsar magnetosphere
the  conditions change with the radius. A wave  which is excited at the
frequency $\omega$ at radius $R_0$ propagates outward into the lower
density plasma. A wave initially at resonance does not remain resonant
as it the propagates. } The frequency width of the resonance is
$\sim\Gamma_r\ll\omega_r$, where $\Gamma_r\propto\omega_r\propto
n^{1/2}\propto R^{-3/2}$. For a given $\omega$ the resonant condition
$|\omega-\omega_r(R)|\lesssim \Gamma_r(R)$ is satisfied only for a small
$\Delta R/R \sim\Gamma_r/\omega_r$. As a consequence, the resonant growth
condition can be met only for a short time, and effectively only at a single
height in the magnetosphere, { which places a severe restriction on the
gain factor $G=\exp2\int\Gamma dR$. It is the gain  factor $G$ which
determines the efficiency of the wave generation and  not the local growth
rate $\Gamma$.} It was for this reason that a pessimistic view of the
effectiveness of the resonant instability was taken in
\cite{MG99}. However, a wave at a given $\omega$ can grow nonresonantly for
$\omega\lesssim \omega_r$, { and the slightly lower growth rate for the
nonresonant instability, compared with the resonant instability, is
relatively unimportant compared with the much greater distance over which
nonresonant growth occurs. The much longer growth path through the
magnetosphere results in a much larger gain factor.  We note  that
during the propagation the wavevector, in general, deviates from the
initial propagation direction, so that the propagation angle
$\theta$ changes, and in principle this would limit the growth if the wave
moves out of resonance as $\theta$ increases. This issue will be 
studied elsewhere.    A wave that
starts growing nonresonantly with $\omega\ll\omega_r$ at some radius,
$R_0$, keeps growing  while propagating outward until
$\omega_r\propto  R^{-3/2}$ decreases to $\omega_r=\omega$. At this resonant
point the beam mode joins on to the L-O mode \citep{Ishimaru}. 
Beyond this point
amplification ceases and the wave escapes as in the L-O mode. }

Let us estimate the gain factor using the approximation for $\Gamma_n$
for the cold plasma case, assuming $\gamma_p$ and $\gamma_b$ do not change
during the outflow, with $\omega_r\propto n_p^{1/2}\propto R^{-3/2}$
\cite{MT77}.   The gain factor at a given $\omega$ is
$G=\exp(2\int_{R_0}^\infty \Gamma(\omega,R) dR)
=\exp [2(\omega_{b0}R_0/\gamma_b^{3/2})x(x^{-2/3}-1)]$, where $\omega_{p0}$
is the plasma frequency at $R=R_0$, and
$x=\omega/2\omega_{p0}\gamma_p^{1/2}$. The factor $G$ is a maximum at
$x\approx 0.2$  and, reverting to ordinary units,
$G_{\rm max}\approx \exp(0.25\omega_{b0}R_0/c\gamma_b^{3/2})$.
{ For numerical estimates we use the following parameters: $P=1\,$s,
$B=10^{12}\,$G, $n_p=Mn_{GJ}=10^2\times 6\cdot 10^{10}\,$cm$^{-3}$ (near
the pulsar surface), $\gamma_p=10$ \citep{DH82,HA01}, $\gamma_b=10^3$,
$n_b/n_p\sim 1$, and  $R_0=10R_p$ \citep{Lyu00}.  This immediately gives that
the maximum is  achieved at $\omega/2\pi\approx 500\,$MHz. The corresponding
gain  factor $\exp(G)\approx \exp(30)\approx 10^{12}$,} which implies
efficient growth. { Lower multiplicities would result in lower
maximum gain frequencies. Lower $\gamma_b$, on the other hand, would
make the instability more efficient. Better knowledge of the pulsar
plasma parameters is required for direct comparison of our predictions
with observations.}  The above estimate is invalid for low frequencies,
corresponding to wave excitation at very large radii $R>10^2 R_0$. At such
radii one has $\omega_r\sim\Omega/\gamma_p$ \cite{Gedalin98}, where $\Omega$
is the gyrofrequency, and then the infinite magnetic field approximation is
no longer valid. Thus there should be a significant decrease of efficiency
of wave growth at low frequencies. Specifically, using parameters chosen by
Ref.~\onlinecite{Gedalin98}, the spectrum should be cut off for
$\omega/\omega_r<10^{-3}$.

Let us summarize the predictions of the proposed  model. Quasi-transverse
waves are generated efficiently in a wide frequency range below
$\omega_{p0}\gamma_p^{1/2}$ for small angles of propagation. The power
spectrum of the escaping radiation is assumed $\propto G^2$, where $G$ is the
gain factor, and there is a maximum in $G^2$ at the frequency
$\omega_{\rm max}\sim 0.1 \omega_{p0}\gamma_p^{1/2}$. The efficiency of wave
growth decreases at both $\omega<\omega_{\rm max}$ and $\omega>\omega_{\rm
max}$, and the rate of decrease (steepness of the curve $G(\omega)$)
increases with the distance from $\omega_{\rm max}$. The
implied form of spectrum is consistent with observations (e.g.,
Ref.~\onlinecite{MT77}).

We conclude that the nonresonant beam instability efficiently generates
quasi-transverse waves in the radio range in a one-step process.  The
growing waves are in a beam mode well below the resonant frequency. As these
waves propagate outward through the magnetosphere they evolve into the L-O
mode when their frequency matches the resonant frequency. It is also
the  place where the growth ceases and, therefore, the spectrum
forms. This emission mechanism  plausibly reproduces the basic features of
the observed radio spectra of pulsar, notably the existence of a maximum
frequency with the spectrum steepening towards both higher and lower
frequencies. { However, a detailed interpretation of the observed power
spectra requires that the emission mechanism proposed here be complemented
with a statistical model for the emission. The observed spectra are obtained
by integration over many pulses, and each pulse probably involves emission
from a statistically large number of individual beams. Moreover, there
remains a potential problem in explaining the lowest frequencies inferred
for some pulsar emission: although the nonresonant model allows emission
considerably below the resonant frequency, a more detailed investigation is
required to determine whether the frequency range can extend low enough. A
more detailed model for pulsar radio emission based on the mechanism
proposed here will be presented elsewhere.}

\acknowledgments
M.G. is grateful to V.V. Usov and Yu.E. Lyubarsky for useful
discussions, and we thank Q. Luo for helpful comments. This research was
supported in part by Israel Science
   Foundation under grant No.  170/00-1.

\end{document}